\documentclass{elsart}
\usepackage{graphicx,amssymb}

\begin{document}

\begin{frontmatter}
\title{Nonextensivity at the edge of chaos of a new universality class of one-dimensional unimodal dissipative maps}
\author{Guiomar Ruiz}
\ead{guiomar.ruiz@upm.es}
 \address{Centro Brasileiro de Pesquisas Fisicas
\\
Rua Xavier Sigaud 150, 22290-180 Rio de Janeiro -- RJ, Brazil \\
and\\
Depto. Matem\'atica Aplicada y   Estad\'\i stica, Universidad Polit\'ecnica de Madrid\\
Pza. Cardenal Cisneros n.4, E-28040 Madrid, Spain}
  \author{Constantino Tsallis}
\ead{tsallis@cbpf.br}
\address{Centro Brasileiro de Pesquisas Fisicas
 and National Institute of Science and Technology for Complex Systems\\
Rua Xavier Sigaud 150, 22290-180 Rio de Janeiro -- RJ, Brazil\\
and\\
Santa Fe Institute, 1399 Hyde Park Road, \\
Santa Fe, NM 87501, USA}
 \date{\today}
\begin{abstract}
We introduce a new universality class of one-dimensional unimodal dissipative maps. The new family, from now on referred to as the ($z_1,z_2$)-{\it logarithmic map},
corresponds to a generalization of the $z$-logistic map. The 
Feigenbaum-like constants of these maps are determined.
It has been recently shown that the probability density of sums of iterates at the edge of chaos of the $z$-logistic map is numerically consistent with a $q$-Gaussian, the distribution which, under appropriate constraints, optimizes the nonadditive entropy $S_q$. We focus here on the presently generalized maps to check whether they constitute a new universality class with regard to $q$-Gaussian attractor distributions.
We also study the generalized $q$-entropy production per unit time on the new unimodal dissipative maps, both for strong 
and weak 
chaotic cases.  The $q$-sensitivity indices are obtained as well.
Our results are, like those for the $z$-logistic maps, numerically compatible with the $q$-generalization of a Pesin-like identity for ensemble averages.
\end{abstract}
\end{frontmatter}
\section{Introduction}
\label{intro}
One-dimensional nonlinear maps play an important role in the development of the theory of chaos. Their long-time behavior is different for different kinds of maps and, for dissipative dynamical systems, the  phase space measure is not conserved: all trajectories approach a certain subset of the phase space called {\it attractor}. The characterization of chaotic attractors is interesting and, as we shall see, a still open problem whenever the Lyapunov exponent vanishes (frequently referred to as {\it weak chaos}).

The exploration of their special dynamical properties is, besides their simplicity and convenience for the development of theory of chaos, also motivated by the hope that the study of the possible limits of validity of the canonical statistical mechanics can benefit from the study of much simpler dynamical systems that are known to exhibit statistical-mechanical analogies \cite{beck}.

In particular, one-dimensional unimodal maps may depend of a single control parameter that determines the dynamical behavior of the map. They typically have only one attractor, which differs for different parameter values. This fact makes these maps to constitute paradigmatic models in the study of the emergence of complexity in dynamical systems.

Here we introduce and analyze a new universality class of one-dimensional unimodal dissipative maps. Our initial scope is to test, through them, the applicability and usefulness of generalized dynamical indicators ($q$-indices) that emerge within nonextensive statistical mechanics \cite{extensive,qsensib} in order to establish a more complete classification of (weak and strong) chaotic systems.

The paper is organized as follows. In Sect.~\ref{sec:1} we introduce the ($z_1, z_2$)-logarithmic maps and we compare their attractors with those corresponding to two well known one-dimensional unimodal dissipative maps. In Sect.~\ref{sec:2} we briefly review the generalized properties we are interested to test, as well as some peculiarities of their numerical study. In Sect.~\ref{sec:3} we present our numerical results. Our main conclusions are drawn in Sect.~\ref{sec:4}.
\section{One-dimensional unimodal dissipative maps: a new class.}
\label{sec:1}
The well known $z$-logistic maps are among the simplest one-dimensional nonlinear dynamical systems that allow a close investigation of  complex behavior. This family reads
\begin{equation}
\label{zlogistico}
x_{t+1}=1-\mu |x_t|^z \, \, \, (z \ge 1; \mu \in [0,2]; |x_t|\le 1)\, ,
\end{equation}
where $\mu$ is the control parameter, whose values are limited in order to avoid the orbits to escape to infinity; $z$ characterizes the inflection of the map in the neighborhood of the extremal point $x=0$. The larger is $z$, the flatter is the $x=0$ maximum. The $z=2$ map is, as is well known, isomorphic to $y_{t+1}\propto y_t (1-y_t)$. These maps are known to have topological properties that do not dependent on $z$, and they constitute important universality classes of unimodal maps. However, their metrical properties, such as Lyapunov exponents, chaos threshold control parameter values and Feigenbaum-like constants, do depend on $z$.  In particular, the control parameter critical value $\mu_c$ monotonically increases from $1$ to $2$, when $z$ increases from $1$ to $\infty$ (we focus here, and hereafter, to the {\it first} entrance to chaos while increasing $\mu$ above zero).

The $z$-exponential family of maps was introduced  \cite{zexpon} to characterize a further degree of flatness that $z$-logistic maps cannot attain [even for $z\rightarrow \infty$ in Eq. (\ref{zlogistico})]. They are inspired in Cauchy's exponential function (infinitely differentiable at $x=0$ and nevertheless nonanalytic). This family is defined as follows:
\begin{equation}
\label{zexponential}
x_{t+1}=1-\mu e^{-1/|x_t|^z} \, \, \, (z>0; \mu \in [0,\mu ^*(z)]; |x_t|\le 1) \,,
\end{equation}
where the upper limit $\mu ^*(z)$ of the control parameter guarantees that orbits do not diverge, and depends slowly from $z$ [e.g., $\mu^*(0.5) \approx 5.43$]. It has been observed that there is  a value of $z$ above which the corresponding attractors are topologically isomorphic to those of the logistic map. These maps enabled to  investigate chaos in a new universality class of maps \cite{zexpon}. However, their extreme flatness caused serious numerical problems when we tried to study various dynamical properties and, very particularly, the probability density of sums of iterates at the edge of chaos \cite{queiros}.

We then propose a new family of maps, to characterize chaotic behavior of an universality class different from that of unimodal maps. From now on we call them {\it $(z_1,z_2)$-logarithmic maps}, and they are defined as follows:
\begin{eqnarray}
\label{zlogarith}
&& x_{t+1}=1-\mu \frac{|x_t|^{z_1}}{\ln^{z_2}{\left(\frac{|x_t|+1}{|x_t|}\right)}}  \nonumber\\
&&(z_1\ge 1; z_2\ge 0;
\mu \in [0,\mu ^*(z_1,z_2)]; |x_t|\le 1)
\end{eqnarray}
where the parameters $(z_1, z_2)$ characterize the map. Notice that $(z_1,z_2)=(1,0)$ ({\it tent map}) makes the map to be not differentiable at $x=0$.
The upper limit $\mu ^*(z_1,z_2)$ varies slowly with $(z_1, z_2)$. These maps generalize the $z$-logistic maps (which are recovered for $z_2=0$). At their extremum, they are {\it less flat} than the corresponding $z_1$-logistic maps, in contrast with the $z$-exponential maps.

\begin{figure}[h]
\begin{center}
\includegraphics[width=14cm,angle=0]{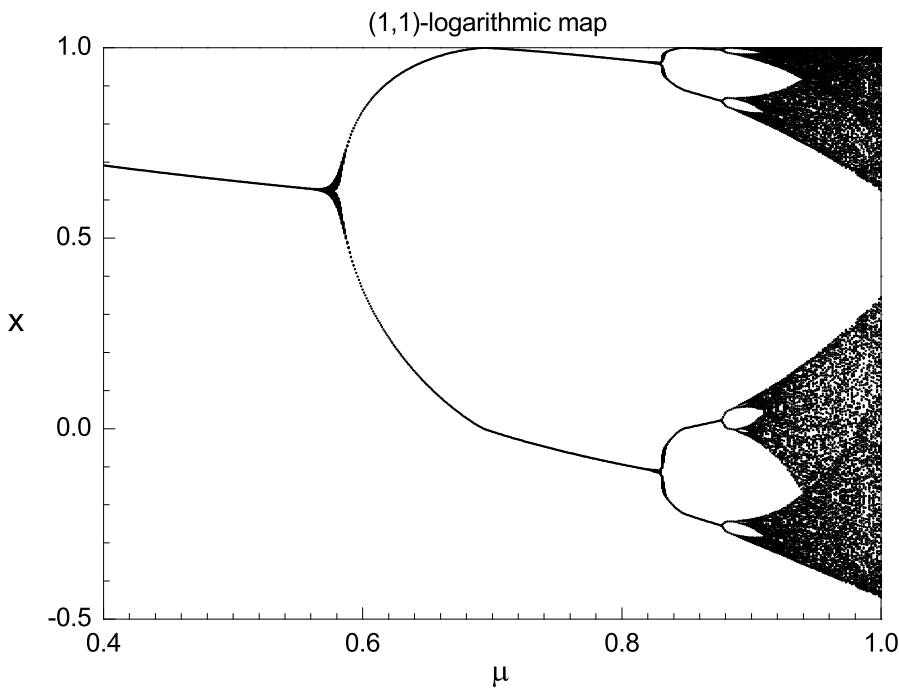}
\end{center}
\caption{Bifurcation diagram of the ($1$,$1$)-logarithmic map.}
\label{fig:1}       
\end{figure}

\begin{figure}[h]
\begin{center}
\vspace{1cm}
\includegraphics[width=6cm,angle=0]{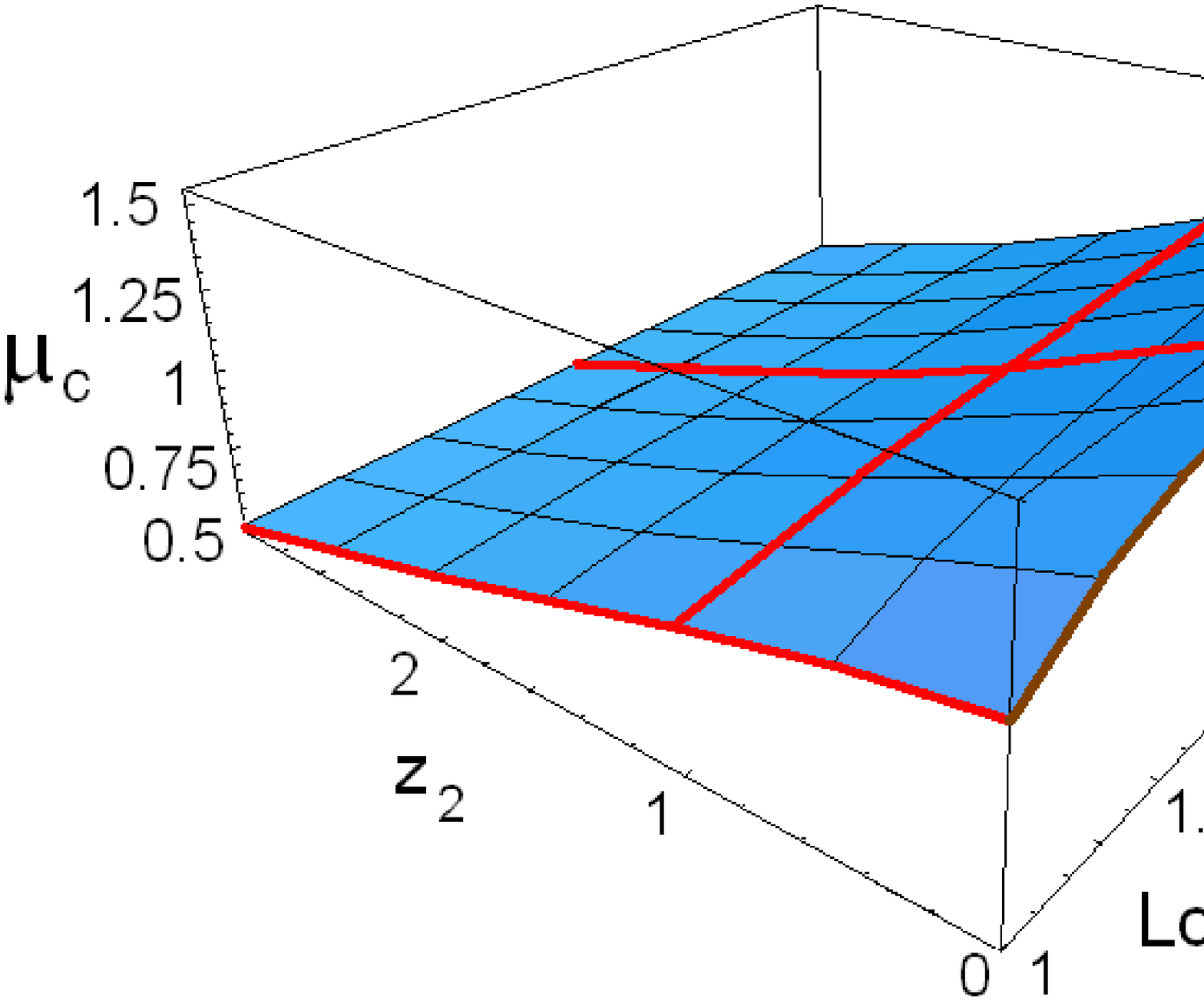}
\hspace{1cm}
\includegraphics[width=6cm,angle=0]{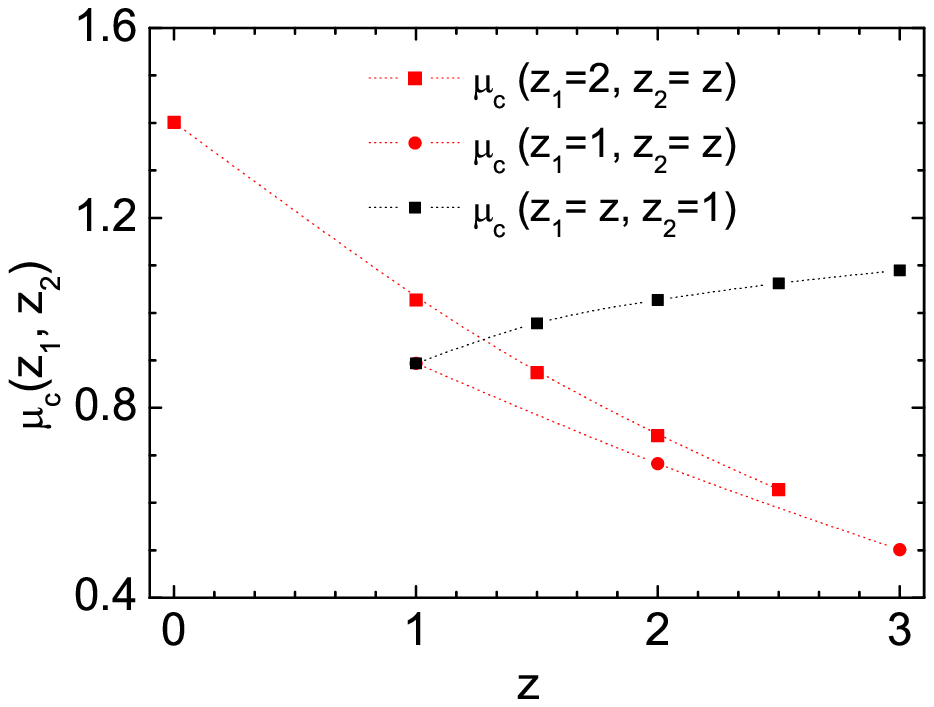}
\end{center}
\caption{\small ($z_1$,$z_2$) dependence of $\mu_c$. The red lines on the surface are also represented on a two-dimensional plot. The brown line represents the critical value $\mu _c(z)$ of the $z$-logistic map.}
\label{fig:2}     
\vspace{2cm}
\end{figure}

We have numerically verified, for a wide range of values of $(z_1, z_2)$, that the attractors of the $(z_1, z_2)$-logarithmic map appear to be topologically isomorphic to those of the logistic map.
As an example, Fig.~\ref{fig:1} exhibits the $\mu$-dependence of the dynamical attractor of the $(1,1)$-logarithmic map.
This bifurcation diagram shows a first whole cascade of period doubling of non chaotic attractors $2^{k-1}\rightarrow 2^k$ at parameter values $\mu _k$, so that the sequence $\{\mu _k\}\;(k=1,2,\cdots)$ approaches the critical value $\mu _c \equiv \mu_{\infty} =0.893425\dots $, the point that defines the chaos threshold. By further increasing $\mu$ above $\mu _c$, the behavior becomes complex and trajectories become chaotic with the exception of those `windows' where  stable periods re-emerge  (see first window at $\mu=0.981\dots $). The entrances to such windows exhibit the {\it intermittency } phenomenon related to a {\it tangent bifurcation}, which implies the existence of a `laminar phase'. Besides the period doubling scenario, the $(z_1,z_2)$-logarithmic maps also exhibit {\it band splitting} phenomenon, so that when approaching $\mu _c$ from the chaotic regime, a chaotic attractor band splits into two
chaotic bands in such a way that the iterates alternate between both bands in a periodic way, even though the movement is chaotic inside each band. A whole cascade of parameter values $\{\hat{\mu }_k\}\;(k=1,2,\cdots)$ exists where there is a splitting from $2^{k-1}$ to $2^k$ chaotic bands. Both the sequences of period doubling parameter values and the band splitting parameter values converge to the critical point $\mu _c$ according to the equation:
\begin{equation}
\label{eqFeigem}
\lim_{k\rightarrow \infty}\frac{\mu _k-\mu _{k-1}}{\mu _{k+1}-\mu _{k}}=\delta(z_1,z_2) \,,
\end{equation}
where $\delta(z_1,z_2)$ is the Feigenbaum-like constant of the $(z_1,z_2)$-logarithmic map: see Table~\ref{tab:1}. We obtain analogous results for other  ($z_1$,$z_2$) values. Therefore we verify that, in spite of the fact that the topological properties do not depend on ($z_1$,$z_2$), metrical properties do. The dependance of $\mu_c$ on $(z_1,z_2)$ is depicted in Fig.~\ref{fig:2}
\begin{table}
\begin{center}
\caption{Numerical values of the critical control parameter $\mu_c(z_1,z_2)$, and of the  Feigenbaum-like constant $\delta(z_1,z_2)$. In some cases, the period doubling sequence convergence is very slow.}
\label{tab:1}       
\begin{tabular}{lll}
\hline\noalign{\smallskip}
($z_1$,$z_2$) & $\mu _{c}$ & $\delta$  \\
\noalign{\smallskip}\hline\noalign{\smallskip}
(1,1)&0.893425\dots &2.9\dots\\
(1,2)&0.68249659\dots &3.09\dots\\
(1,3)&0.5009906118\dots &3.24\dots\\
(2,0)&1.401155189\dots &0.66\dots\\
(2,1)&1.027082958927880\dots &4.775\dots\\
(2,1.5)&0.8739420229318233\dots &4.83\dots \\
(2,2)&0.7413182584853554\dots &4.89\dots\\
(2,2.5)&0.6272310052913988\dots &4.95\dots\\
(1.25,1)& 0.94343421 \dots &3.45\dots \\
(1.3,1)&0.951193254 \dots &3.54\dots \\
(1.4,1)& 0.9653357702\dots &3.74\dots \\
(1.5,1)&0.9779976050295\dots &3.91\dots \\
(1.6,1)&0.98947839547\dots &4.12\dots \\
(1.75,1)& 1.004931008022 \dots &4.38\dots \\
(2.5,1)&1.06221102203249533\dots &5.52\dots\\
(3,1)&1.089407442252358479\dots &6.19\dots\\
\noalign{\smallskip}\hline
\end{tabular}
\end{center}
\end{table}

\section{Generalized properties of chaotic systems}
\label{sec:2}
We are now ready to characterize chaotic attractors of the new family of maps. The $z$-logistic maps, and many others, have already been deeply studied \cite{zexpon,robledo,ugur1,ugur2}. However, some questions remain still open, especially at the edge of chaos.
Furthermore, some already known properties might be tested on the presently introduced new family of nonlinear dynamical systems.

In order to characterize chaos, we may say that a dynamical system is `chaotic' if it possesses sensitive dependence on the initial conditions. That means that randomly chosen very close initial conditions generate totally different trajectories in the long-time run. When the temporal discrepancy increase is exponential, the system is said to be {\it strongly chaotic} (or just {\it chaotic}). But it is known that there are ubiquitous natural and artificial dynamical systems, typically at the edge of chaos, that exhibit a power-law sensitivity to the initial conditions. Such systems are termed  {\it weakly chaotic} ones. In this case, standard dynamical indicators give a poor description of the complexity of their time evolution. In particular, the Lyapunov exponent vanishes. Nonextensive statistical mechanics  \cite{qsensib} provides a generalization of the standard dynamical indicators of chaos to characterize, not only strongly, but also weakly chaotic systems.

Consistently, the first property we study is the sensitivity to initial conditions. The second property is the $q$-generalized entropy production per unit time (a concept that is devised to be analogous to the Kolmogorov-Sinai entropy rate for strongly chaotic systems). In this work we are interested in studying these two properties on averages (denoted as $\langle \cdots \rangle$) over the entire phase space of the maps ($-1\le x\le 1$) instead of using the quickest-spreading special regions \cite{latora}. This kind of averaging mimics Gibbs' approach to thermostatistical ensembles, and minimizes the role of fluctuations (particularly large at the edge of chaos).
The third property we are also interested in is the characterization of the probability density of sums of iterates of the map and, consequently, the possible applicability of a $q$-generalized  Central Limit Theorem (CLT) to the iterates of {\it deterministic} dynamical systems.

It has been shown in many one-dimensional unimodal maps that the sensitivity to initial conditions is given by the generalized functional expression   \cite{qsensib} (called $q$-exponential function within the context of nonextensive statistical mechanics)
\begin{equation}
\label{sensitivity}
\xi(t) \equiv \lim _{\Delta x(0)\rightarrow 0} \frac{|\Delta x(t)|}{|\Delta x(0)|}= [1+(1-q)\lambda _q t]^{1/(1-q)}\equiv e_q^{\lambda _q t} \,,
\end{equation}
where $\Delta x(t)$ is the temporal dependence of the discrepancy of two very close initial conditions at time $t$, $q$ is a parameter (for $q=1$ the expression recovers the standard exponential dependence $e^{\lambda_1 \,t}$), and $\lambda_q$ is a generalized Lyapunov coefficient (when $q=1$ then $\lambda _1=\lambda$ and the Lyapunov exponent is recovered; when $q<1$ and $\lambda _q>0$ the system is said to be weakly sensitive to the initial conditions; when  $q>1$ and $\lambda _q<0$ it is said to be weakly insensitive).

We consider very close initial conditions, randomly chosen within the interval $[-1,1]$, from which we estimate  $\xi(t)$.
We perform this operation many times (typically $10^7$), and average all the values of the corresponding $\ln _q \xi(t)$ (where $\ln _q x\equiv (x^{1-q}-1)/(1-q)$ is the inverse function of the $q$-exponential; $\ln_1 x =\ln x$) for various values of $q$. We consider increasingly small initial discrepancies $\Delta x(0)$ between each pair of initial conditions such as to obtain results which no further depend on the value of $\Delta x(0)$ for increasingly long times. We vary the value of $q$ and verify a nontrivial property \cite{zexpon}, namely that a special value of $q$, noted $q_{sen}^{av}$ (where {\it sen} and {\it av} stand respectively for {\it sensitivity} and {\it average}), exists which yields  a {\it linear} dependence with time. In other words, we verify that $\langle \ln _q \xi \rangle (t)\simeq \lambda _{q_{sen}^{av}}^{av} t$, where the linear coefficient $\lambda _{q_{sen}^{av}}^{av}$ constitutes a $q$-generalized Lyapunov coefficient.

With respect to the entropy production per unit time,  it is known that the Boltzmann-Gibbs entropy ($S_{BG}\equiv -\sum _{i=1}^{W}p_i \ln{p_i}$) is the appropriate one when strong chaos is present. But, at the edge of chaos (see  \cite{extensive,qsensib,latora} and references therein), we may conveniently use the $q$-generalized entropy
\begin{equation}
\label{entropy}
S_q=\frac{1-\sum _{i=1}^{W}p_i^q}{q-1} \;\;\;(S_1=S_{BG}) \,.
\end{equation}
For each universality class of maps (characterized by $(z_1,z_2)$), one special value of the entropic index $q$, noted $q_{ent}$ (where {\it ent} stands for {\it entropy}), exists for which the entropy production is {\it finite}. We expect, at the light of many maps that have been previously studied, this entropic index to coincide with the one obtained from the study of the sensitivity to initial conditions (Eq.  \ref{sensitivity}), i.e.,
$q_{sen}=q_{ent}$. Other methods (e.g., based on multifractality) do exist for  the calculation of $q$, but we address here just these two ones.

The estimation of the $q$-entropy production  consists in dividing the phase space $x$ in $W$ (typically $10^5$) equal cells, and putting $N_{ic}\gg W$ randomly chosen initial conditions inside one of the $W$ cells. We accompany the spread of points within the phase space, and calculate $S_q(t)$ from the set of occupancy probabilities $\{p_i(t)\}\;(i=1,2,\cdots,W)$. We repeat the operation many times (typically $10^3-10^4$ for strong chaos, and $W/2$ for weak chaos), choosing different initial cells within which the $N_{ic}$ initial conditions are chosen (we usually use $N_{ic}=10\,W$). Finally, we average the entropies $S_q(t)$ over the $N_c$ initial cells so that the proper value of the entropic parameter $q_{ent}^{av}$ is the special value of $q$ which makes the averaged $q$-entropy  $\langle S_{q_{ent}^{av}} \rangle_{N_c}$ production to be {\it finite}. The $q$-entropy production per unit time
\begin{equation}
\label{KSentropy}
K_{q_{ent}^{av}}\equiv \lim _{t\rightarrow \infty} \lim _{W\rightarrow \infty} \lim _{N_{ic}\rightarrow \infty}\frac{\langle S_{q_{ent}^{av}} \rangle _{N_c}}{t}
\end{equation}
is calculated taking into account that the partitions of phase space must be such as to obtain robust results.

We also investigate the probability density of sums of iterates of the maps. The iterates of a deterministic dynamical system can never be completely independent, since they are generated by a deterministic algorithm. However,  a Central Limit Theorem (CTL) for {\it deterministic} dynamic systems  can be proved \cite{CLT} when we consider a one-dimensional map, $x_{t+1}=f(x_t)$, with positive Lyapunov exponent. More precisely,
the well known CLT assumption about the independence of $N$ identically distributed random variables is replaced by a weaker property that essentially means asymptotic statistical independence for large time difference. In particular, the probability distribution of the rescaled sum
\begin{equation}
\label{distbnorm}
y= \frac{1}{N^{\gamma}}\sum_{t=1}^{N}g(x_t)
\end{equation}
becomes a Gaussian for the number of iterates $N\rightarrow \infty$, regarding the initial value $x_1$ as a random variable ($\gamma=1/2$ for strongly chaotic maps).  Here $g:\Re^d\rightarrow \Re^k$ is a suitable smooth function with vanishing average which projects from the $d$-dimensional phase space to a $k$-dimensional subspace. In our case, $d=k=1$, and $g(x_t)=x_t-\langle x \rangle$. It is rigorously proved that the conditions of validity of a CLT, due to the mixing property associated with strong chaos, are satisfied for the logistic map, $\mu=2$. A CLT has not been rigorously proved neither for other parameter values nor for other $z$-logistic maps, but Gaussian limit behavior is also numerically observed in \cite{ugur1} for other $z$ values on strong chaos regime.  Consistently, we also expect to verify the Gaussian limit behavior on strong chaotic ($z_1, z_2$)-logarithmic maps.
It is clear, however, that this CTL does not hold at the critical points, where the Lyapunov exponent vanishes.
Due to strong correlations between the iterates, a non-Gaussian limit behavior is expected in those points \cite{ugur1,ugur2}.

On the other hand, it is well known that $q$-Gaussian distributions
\begin{equation}
\rho(y) \propto e_q^{-\beta_q y^2}=\left(1+\beta_q \left(q-1\right)y^2\right)^{1/(1-q)}
\end{equation}
maximize the entropy $S_q$ under appropriate constraints. Consequently, on weakly chaotic $z$-logistic maps, where the $q$-generalized entropy (\ref{entropy}) is to be used, numerical indications of a $q$-generalized CLT are available \cite{ugur1,ugur2}. With this respect, we are interested here in studying the distribution of the  rescaled sums of iterates (\ref{distbnorm}) for the ($z_1$,$z_2$)-logarithmic maps. Notice that the rescaling factor $N^\gamma$ can be absorbed by calculating the variance $\sigma$ of the non-rescaled sum ($\gamma=0$ in Eq.(\ref{distbnorm})) for a given $N$, and then plotting the histogram of the variable $y/\sigma$.

The sum (\ref{distbnorm}) must be evaluated for initial conditions located close to the space phase attractor. Consequently, we may omit the first iterates (i.e.,  a {\it transient}) until we obtain transient-independent distributions (a typical length of transient is up to $2^{11}$). This is of course irrelevant in the $N\to\infty$ limit, but it is numerically convenient when we must use finite values of $N$ (typically up to $2^{22}$). We use quadruple precision of Intel Fortran, to avoid roundoff induced effects.

\section{Numerical results}
\label{sec:3}
We studied the sensitivity to initial conditions in both strong and weak chaos regimes.
For strong chaos we obtain $q_{sen}^{av}=q_{ent}^{av}=1$. This means that the Boltzmann-Gibbs entropy  is the appropriate one for this regime. For weak chaos, instead, we obtain  $q_{sen}^{av}=q_{ent}^{av}<1$.

\begin{figure*}
\begin{center}
\includegraphics[width=6.3cm,angle=0]{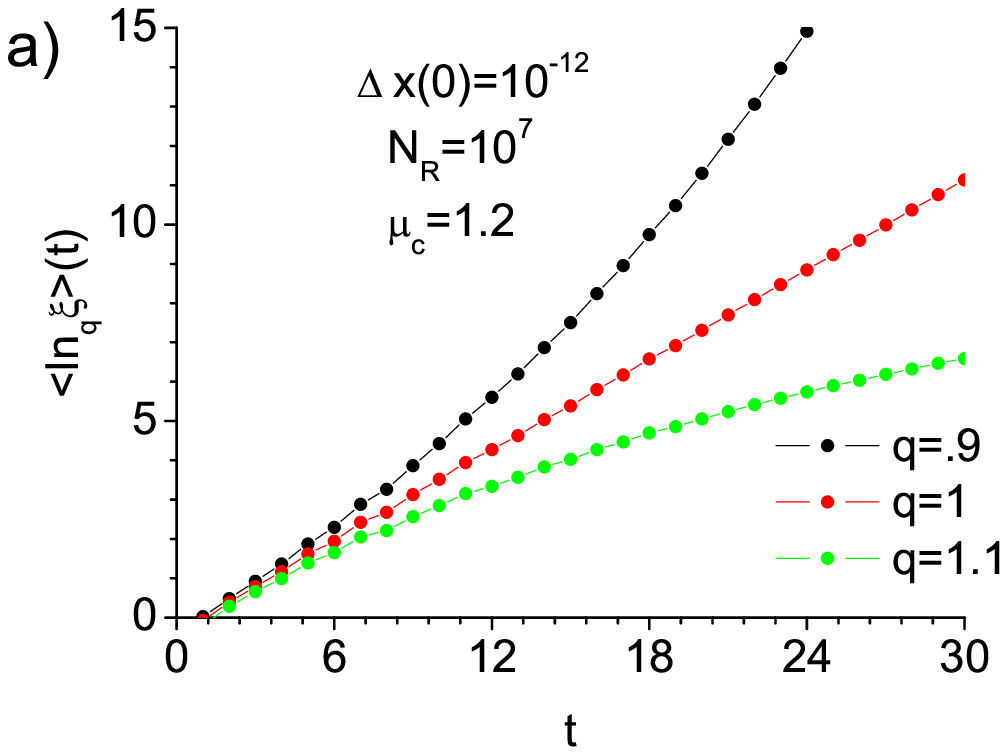}
\includegraphics[width=6.3cm,angle=0]{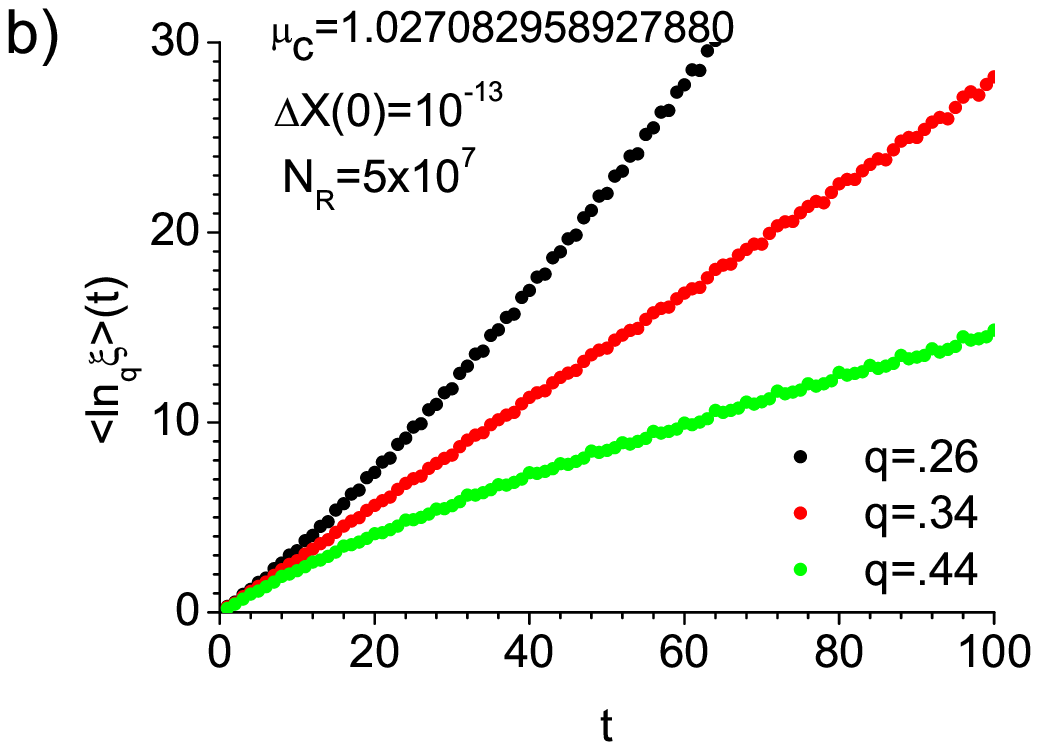}
\includegraphics[width=6.4cm,angle=0]{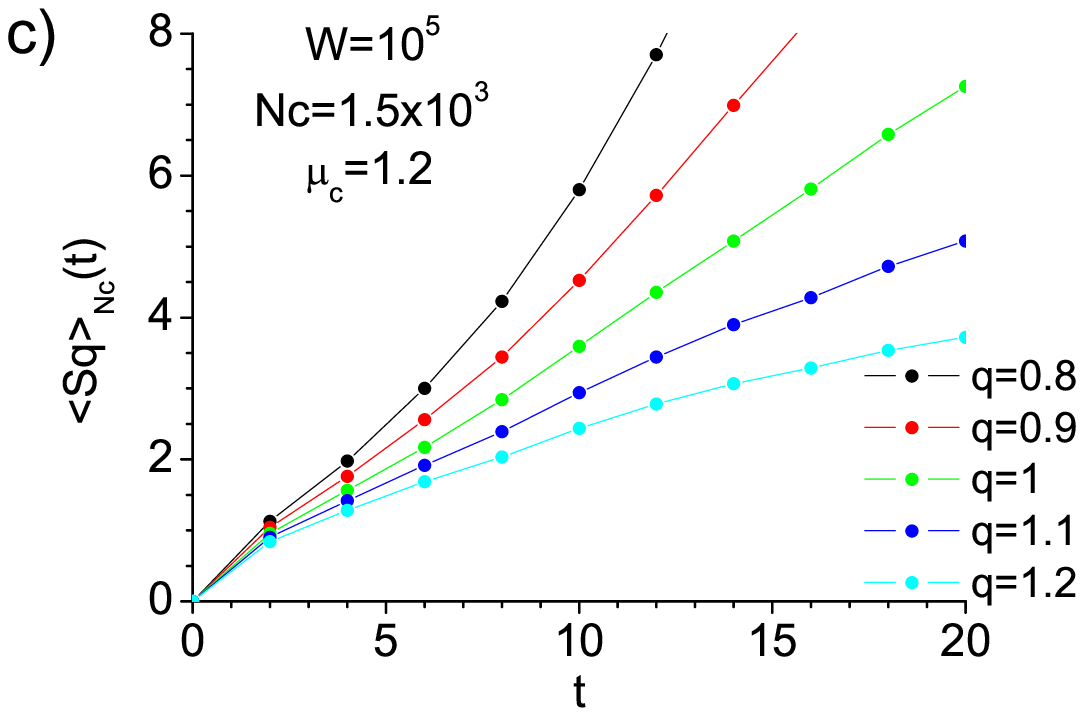}
\includegraphics[width=6.2cm,angle=0]{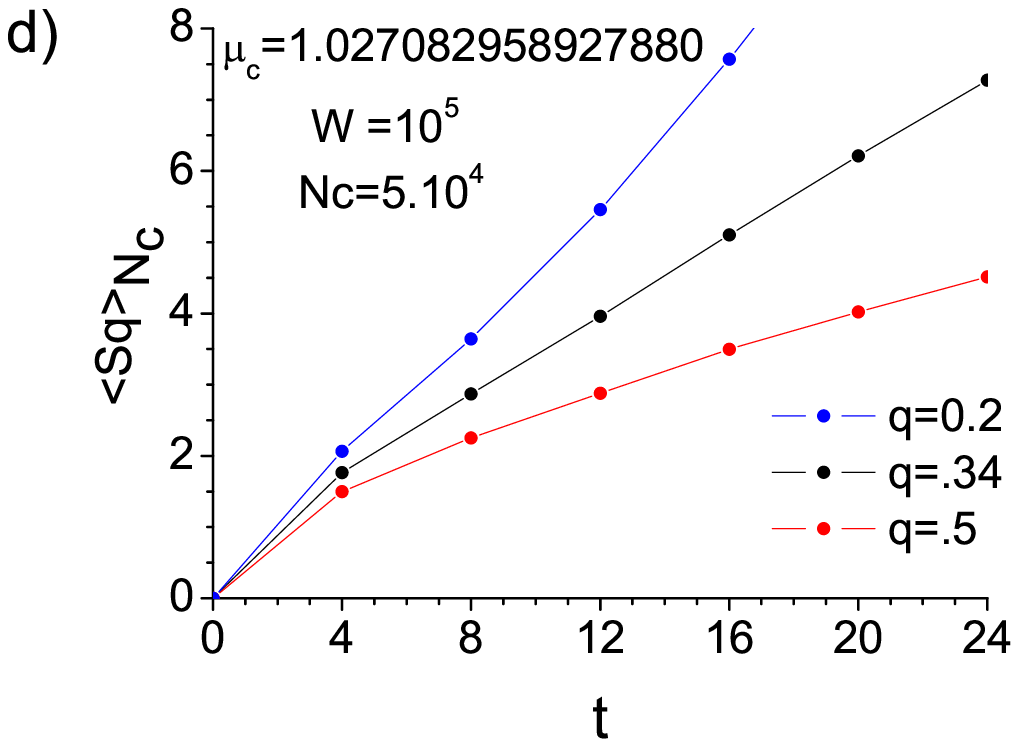}
\includegraphics[width=6.1cm,angle=0]{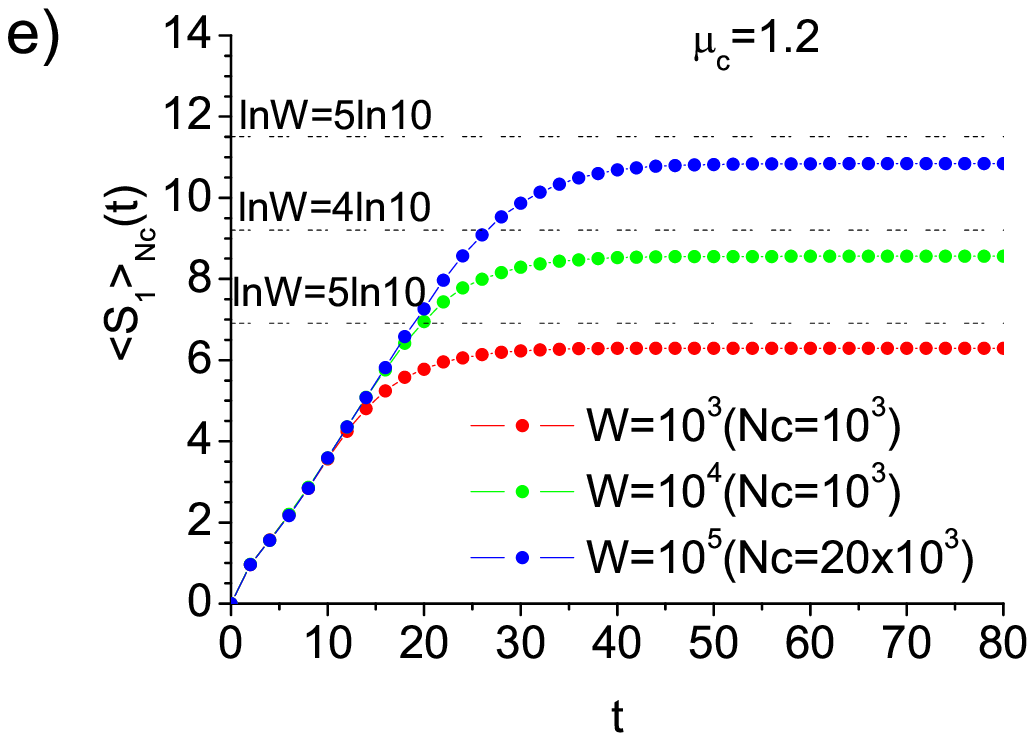}
\includegraphics[width=6.3cm,angle=0]{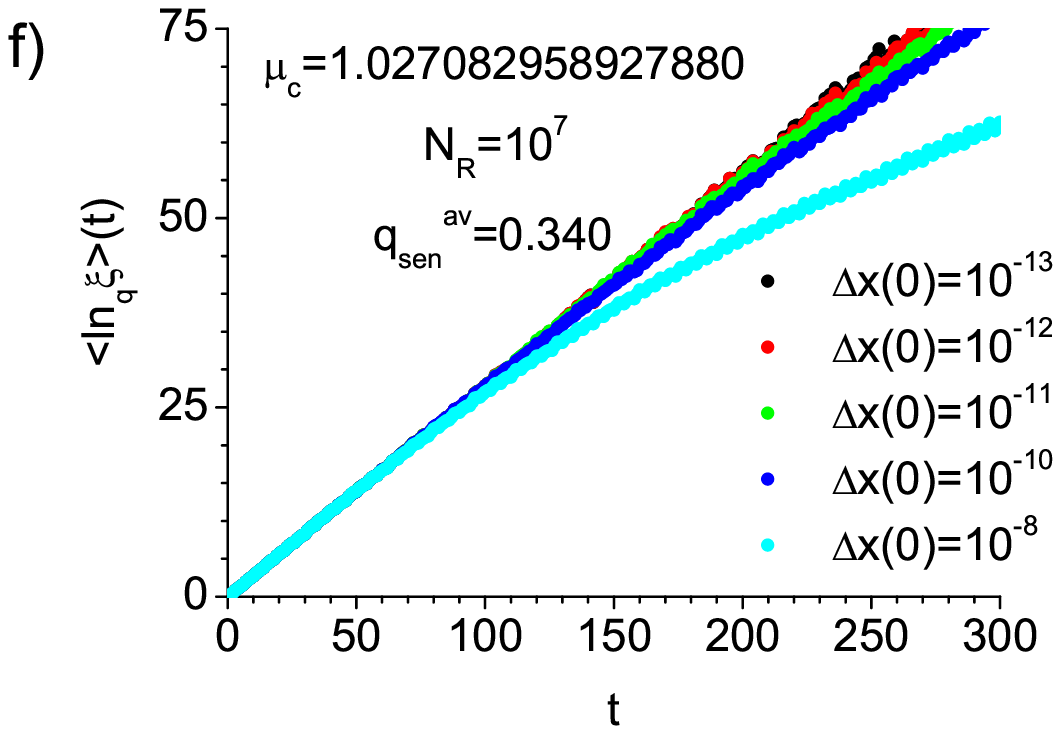}
\end{center}
\caption{\small {$(2,1)$-logarithmic map. Time dependence of the  of $\langle \ln _q \xi \rangle$ (figures (a)-(b)) and  $\langle S_q \rangle$ (figures (c)-(d)) for strong and weak chaos.
The values of the control parameter are $\mu=1.2$ for strong chaos, and $\mu  = 1.027082958927880$ for weak chaos.
Figure (e) shows the convergence of $\langle S_q \rangle$, for strong chaos, to the infinite fine grain limit ($q_{ent}^{av}=1$). The $q$-entropy function  $\langle S_q \rangle$ averages over $N_c=W/2$ cells. See that series of finer partitions in phase space do not change the value of the $q$-entropy production per unit time but,  when increasing $W$, the first stage remains for a longer time: the rate of $q$-entropy production for far-from-equilibrium evolution depends neither on the number $N$ of points of the initial ensemble, nor on $W$. Figure (f) shows the convergence of $\langle \ln _q \xi \rangle$, for weak chaos, to the infinitely small initial discrepancy of two
 different realizations ($q_{sen}^{av}=0.340$).
Sensitivity function $\langle \ln _q \xi \rangle (t)$ averages over $N_R=10^7$ pairs of realizations; unless otherwise indicated, the discrepancy we consider is $\Delta x(0)=10^{-13}$. }}
\label{fig:3}
\end{figure*}

These facts are illustrated in Fig.~\ref{fig:3} for the $(2,1)$-logarithmic map.
We obtained these entropic indices by fitting, over the intermediate regime (occurring before saturation), the curves with the polynomial $A+Bt+Ct^2$ and comparing their nonlinearity measure $R\equiv C/B$. The optimum value of the entropic index corresponds to $R=0$ (a straight line). The intermediate regime that we consider is such that the linear regression coefficient is constant (typically $0.9999$). Intrinsic fluctuations still persist, in spite of averaging (see Fig.3). We overcome them by studying the composed maps $f^{(j)}$ ($j=2,4$).

\begin{figure}[h]
\begin{center}
\vspace{0.5cm}
\includegraphics[width=6.8cm,angle=0]{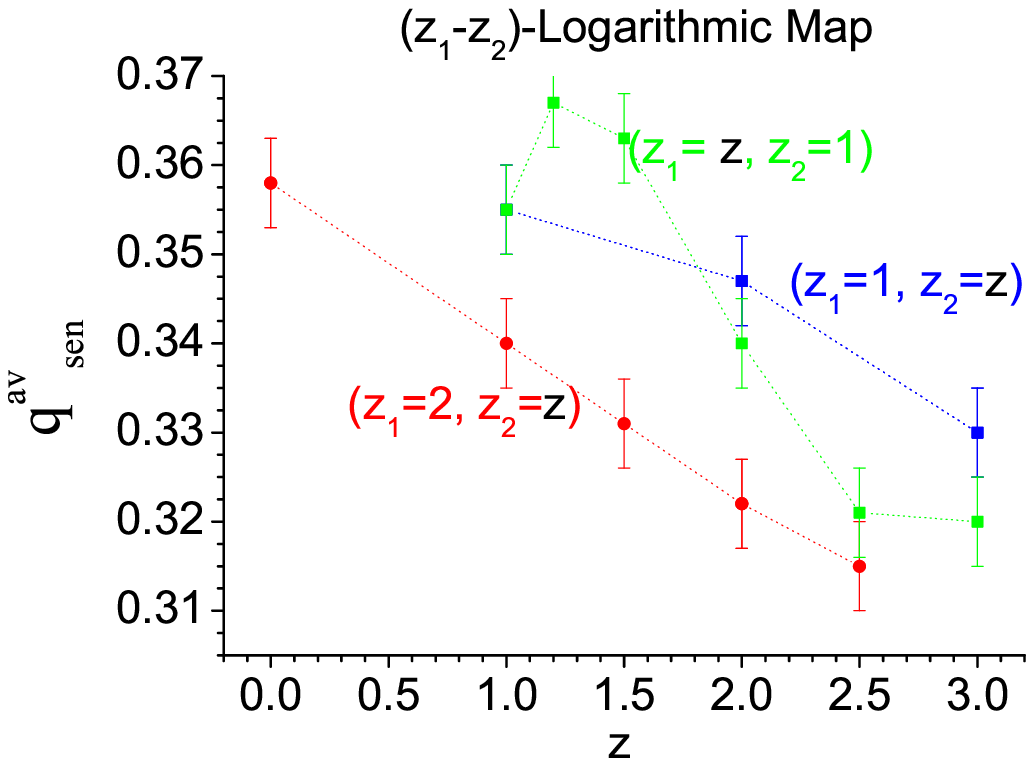}
\includegraphics[width=6.8cm,angle=0]{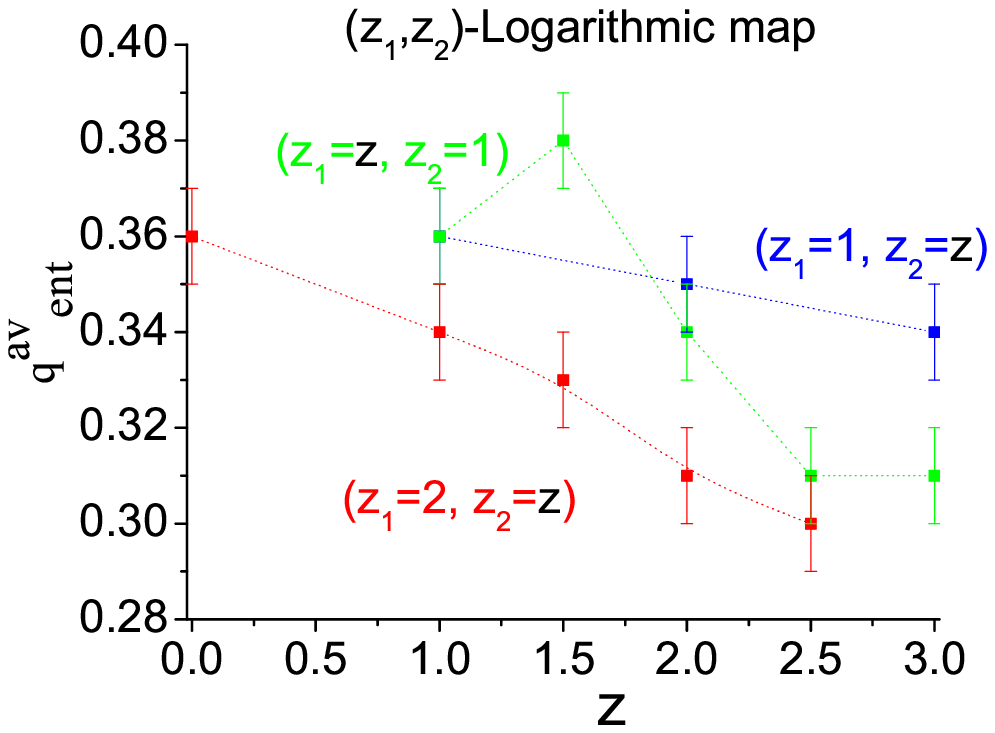}
\end{center}
\caption{\small $(z_1,z_2)$-dependence of $q_{sen}^{av}$ and $q_{ent}^{av}$. These non-monotonic functions coincide within some small error bar. Dotted lines are guides to the eye.}
\label{fig:4}    
\vspace{0.5cm}
\end{figure}

The same procedure is applied to other $(z_1,z_2)$-logarithmic maps (see Table~\ref{tab:2}). In all cases we obtain, within a small error bar, $q_{sen}^{av}=q_{ent}^{av}$.
In Fig.~\ref{fig:4} we can see the influence of $(z_1,z_2)$ on the $q$-indices.
\begin{table}
\begin{center}
\caption{Numerical values, within the $(z_1,z_2)$-logarithmic family of maps, of $q_{sen}^{av}$ and $q_{ent}^{av}$ indexes, on the Feigenbaum attractor (weak chaos).}
\label{tab:2}       
\begin{tabular}{lll}
\hline\noalign{\smallskip}
($z_1$,$z_2$)& $q_{sen}^{av}$ & $q_{ent}^{av}$   \\
\noalign{\smallskip}\hline\noalign{\smallskip}
(1,1)&$0.355\pm 0.005$ & $ 0.36 \pm 0.01$\\ 
(1,2)& $0.347\pm 0.005$& $0.35\pm 0.01$ \\
(1,3)&$0.330\pm 0.005$&  $0.34 \pm 0.01$\\ 
(2,0)&$0.358\pm 0.005$&$0.36 \pm 0.01$\\ 
(2,1)&$0.340\pm 0.005$&$0.34\pm 0.01$\\
(2,1.5)&$0.336\pm 0.005$&$0.33\pm 0.01$ \\ 
(2,2)&$0.322\pm 0.005$&$0.31\pm 0.01$\\ 
(2,2.5)&$0.315\pm 0.005$&$0.30\pm 0.01$\\
(1.5,1)&$0.363\pm 0.005$& $0.38\pm 0.01$\\
(2.5,1)&$0.321\pm 0.005$&$0.30\pm$0.01 \\
(3,1)&$0.300\pm 0.005$ & $0.31\pm 0.01$\\
\noalign{\smallskip}\hline
\end{tabular}
\end{center}
\end{table}

Another interesting result that emerged is the coincidence, for both chaotic regimes, of the slopes of the sensitivity and entropy functions of time. For the $(2,1)$-logarithmic map, we have $K_{1}^{av}=0.382\pm 0.005 \approx \lambda _{1}^{av}=0.372\pm 0.007$ (for strong chaos) and  $K_{0.34}^{av}=0.27\pm 0.01\approx \lambda _{0.34}^{av}=0.28\pm 0.01$ (for weak chaos). These results reinforce those in  \cite{zexpon}, as they are numerically compatible with the $q$-generalized Pesin-like identity for ensemble averages.

\begin{figure}[h]
\begin{center}
\vspace{1cm}
\includegraphics[width=8cm,angle=0]{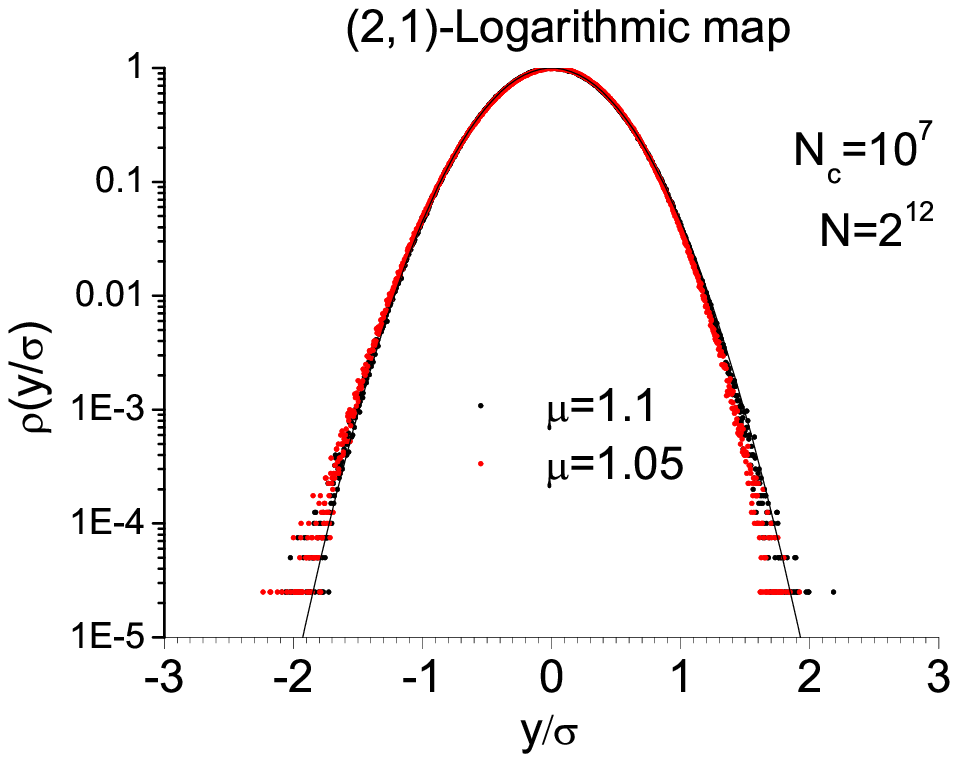}
\end{center}
\caption{\small Numerical estimation of the normalized probability distribution of the  rescaled sums of  iterates of $(2,1)$-logarithmic map, for two different control parameter values ($\sigma=1.85$ for $\mu=1.05$, and $\sigma=6.62$ for $\mu=1.1$). To study the problem near the attractor, we exclude a transient of $N_t=2^{12}$ iterates. $N_c=10^7$ is the number of random initial conditions and $N=2^{12}$ is the number of iterates. Black line corresponds to the Gaussian function $e^{-\beta (y/\sigma)^2}$, $\beta=3.1$.}
\label{fig:5}       
\end{figure}

The probability distribution of the  rescaled sums of iterates of the strongly chaotic $(2,1)$-logarithmic  map presents, as expected, a Gaussian shape (see Fig.~\ref{fig:5}).
The weakly chaotic regime is more subtle. Indeed, it turns out to be necessary to gradually approach the exact value of $\mu_c$ in order to attain the limit distribution.
The critical parameter $\mu$ should approach $\mu _c$, while the number of iterates $N$ of the sum (\ref{distbnorm}) should diverge. In practice, $N$ must be large enough to verify the $N\rightarrow \infty$ TCL assumption, but not so large that the system ``realizes" that $\mu$ is not exactly $\mu _c$.

\begin{figure}[h]
\begin{center}
\includegraphics[width=9cm,angle=0]{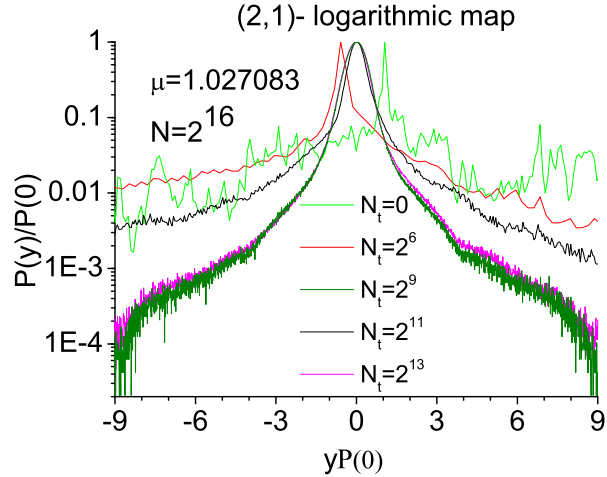}
\end{center}
\caption{\small Numerical estimation of the normalized probability distribution of the rescaled sums of iterates of $(2,1)$-logarithmic map for the control parameter value $\mu = 1.027083$. We  consider different transient times $N_t$. $N$ is the finite number of iterates considered on the sum estimation.}
\label{fig:6}
\vspace{1cm}
\end{figure}

\begin{figure}[h]
\begin{center}
\includegraphics[width=10cm,angle=0]{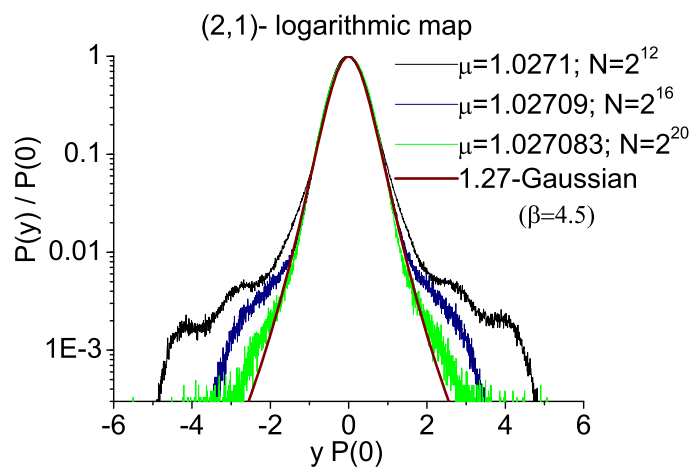}
\end{center}
\caption{\small Data collapse of probability density functions for $\mu \rightarrow \mu _c$, and consistently enlarging $N=2^{2n}$ (see Fig.10). }
\label{fig:7}      
\vspace{0.5cm}
\end{figure}

\begin{figure}[h]
\begin{center}
\includegraphics[width=10cm,angle=0]{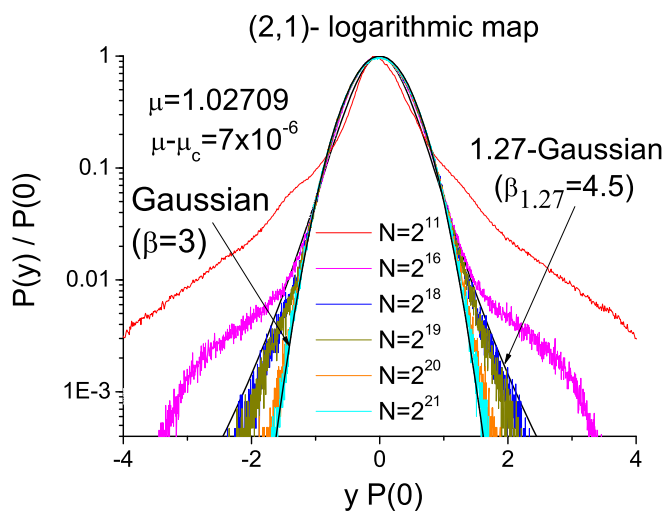}
\end{center}
\caption{\small Rescaled probability density function at the edge of chaos with $4$ digits precision ($\mu=1.02709$). For $N=2^{18}$ it presents a $1.27$-Gaussian behavior in its central part. For $N \gg 2^{19}$, $q\rightarrow 1$, and probability density function tends to a Gaussian.  }
\label{fig:8}       
\end{figure}

\begin{figure}\begin{center}
\includegraphics[width=10cm,angle=0]{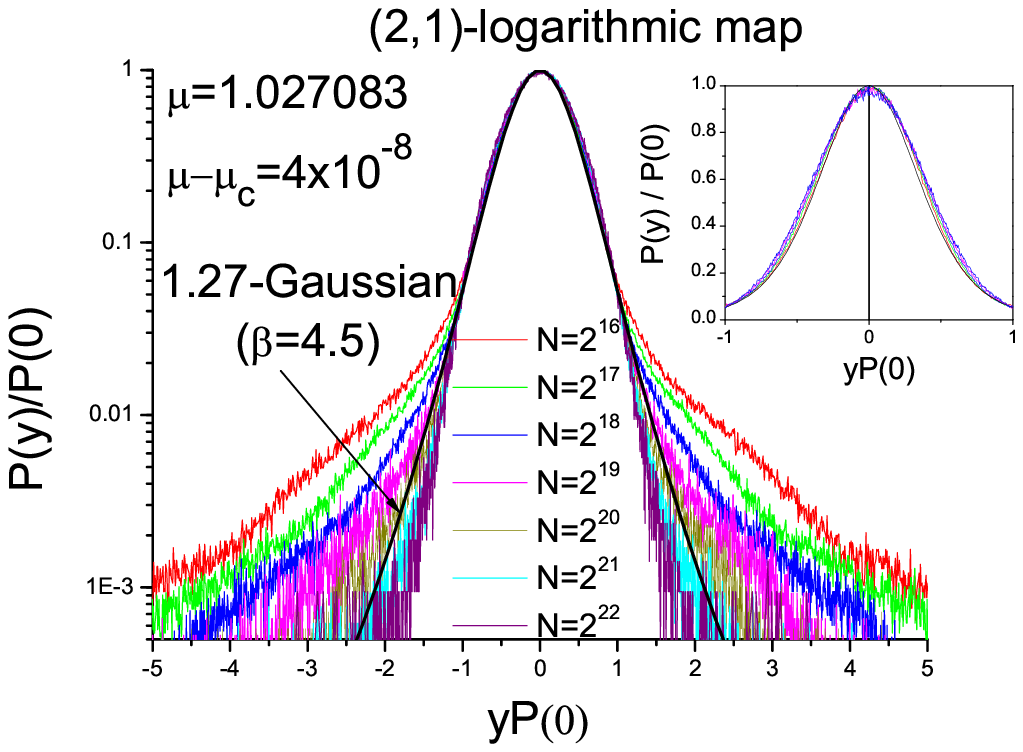}
\end{center}
\caption{\small  Approach to a q-Gaussian distribution with $q=1.27$, for an approximated critical value with a $5$-digit precision $\mu=1.027083$ ($\Delta\mu=4\times 10^-8$).  }
\label{fig:9}       
\end{figure}

First of all, we check that a transient time $N_t=2^{12}$ is enough to consider trajectories close to the phase space attractor. Omitting this transient, the distributions of the sums become independent of the transient length in all cases (see Fig.~\ref{fig:6}). Summarizing, instead of \ref{distbnorm} we use
\begin{equation}
\label{distbnorm2}
y= \frac{1}{N^{\gamma}}\sum_{t=N_t}^{N}g(x_t) \,.
\end{equation}
See, in Fig.~\ref{fig:7}, the probability density functions when we gradually approach to the exact value of $\mu _c$. An optimum intermediate value of $N$ exists, for which the data collapse is produced. To obtain the $1.27$-Gaussian convergence for an even more precise value of $\mu _c$, we expect that a much larger numerical value of $N$ ($N\gg 2^{22}$) is needed. The numerical experiment becomes therefore computationally untractable.
Fig.~\ref{fig:8}--\ref{fig:9} illustrate the effect, on the distribution shape, of considering a finite value of $N$ for two finite-precision values of the critical parameter $\mu_c$.
In both cases, we find that {\it intermediate} values of $N$ show $q$-Gaussian behavior. When $N$ is not large enough (i.e., $N=2^{15}$ for $\mu=1.027083$), the limit distribution exhibits a peaky shape in its central part, due to the fact that summation given by Eq. \ref{distbnorm}  is not adequate to approach the edge-of-chaos limiting distribution.

\begin{figure}
\begin{center}
\includegraphics[width=13cm,angle=0]{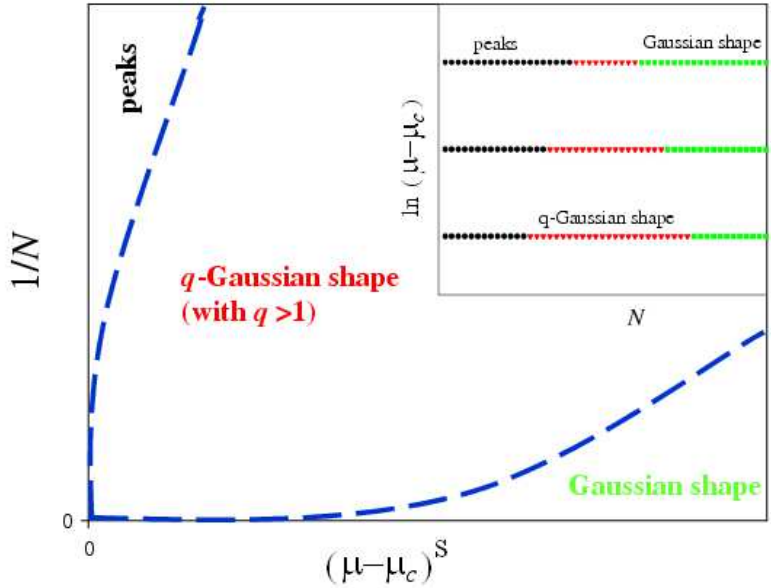}
\end{center}
\caption{\small Shape of the rescaled probability density function of a map at the edge of chaos, for finite $N$ and finite precision for the critical parameter $\mu_c$ ($\mu$ is the control parameter of the map). $\lim_{N \rightarrow \infty} \lim_{\mu \rightarrow \mu_c+0} PDF= peaky$; $\lim_{\mu \rightarrow \mu_c+0} \lim_{N \rightarrow \infty} PDF= Gaussian$; $\lim_{N \rightarrow \infty}PDF= q-Gaussian$ $[fixed \, N (\mu-\mu_c) \in (0, \infty)]$. By courtesy of the authors of Ref \cite{ugur2}, where $\mu \equiv a$, $\mu_c \equiv a_c$, and $s=\log{4}/\log{\delta}$.}
\label{fig:10}       
\end{figure}

Fig.~\ref{fig:10} sketchily  shows the effect, on the shape of the distribution, of considering approximate values of the critical parameter with gradually improved precision, while $N$ increases. We then find a plateau of $N$ values for which finite summation is  adequate to approach the edge-of-chaos limiting distribution, which plausibly is a $q$-Gaussian.

This scenario is fully consistent with the one observed for the $z=2$ logistic map \cite{ugur2}.
\section{Conclusions}
\label{sec:4}
Let us summarize our main results:

(i) A new universality class of one-dimensional unimodal dissipative maps is introduced, characterized by a degree of flatness smaller than that of $z$-logistic maps.

(ii) The critical value on chaos threshold $\mu _c(z_1,z_2)$ and the Feigenbaum-like constant are numerically estimated for $(z_1,z_2)\in[1,3]\times [0,3]$. Topological properties do not depend on $(z_1,z_2)$
 but metrical properties do.

(iii) The entropic index $q_{ent}^{av}$, which makes the average of the $q$-entropy production finite, and $q_{sen}^{av}$ coincide. This result is in accordance with the behavior of other classes of maps. For strong chaos we verify $q_{ent}^{av}=q_{sen}^{av}=1$.

(iv) The $q$-generalization of the Pesin-like identity is verified for ensemble averages for the $(z_1,z_2)$-logarithmic maps, in both strongly and weakly chaotic cases.

(v) Quantitative $(z_1,z_2)$-dependence of the indices $q_{ent}^{av}=q_{sen}^{av}$ is numerically studied for a certain range of $(z_1,z_2)$ values.

(vi) The probability distribution of the sums of iterates in a strongly chaotic  $(z_1,z_2)$-logarithmic map is a Gaussian, as  expected from the Central Limit Theorem for deterministic chaotic systems.

(vii) The probability distribution of the sums of iterates for weakly chaotic $(z_1,z_2)$-logarithmic map (i.e., at the edge of chaos) appears to approach a $q$-Gaussian, the probability distribution that maximizes the nonadditive entropy $S_q$. These numerical results are consistent with a $q$-generalized Central Limit Theorem \cite{qTCL}.

 All these results are expected to contribute to the correct interpretation of various experimental features in dissipative dynamical complex systems   \cite{Dougl,Liu}.
\section*{Acknowledgements}
We thank interesting remarks by Luis G. Moyano, Evaldo M.F. Curado, U. Tirnakli, C. Beck and Miguel Romera.
We acknowledge partial financial support by CNPq, Capes and Faperj (Brazilian Agencies) and DGU-MEC (Spanish Ministry of Education) through Project PHB2007-0095-PC. One of us (G. R.) acknowledges financial support from  the Universidad Polit\'{e}cnica de Madrid.

\end{document}